\documentstyle[aps]{revtex}
\def\ii{\'\i}
\input psfig
\begin{document}
\title{The Forward Proton Detector at D\O}
\author{Gilvan A. Alves$^{\dagger}$\footnote{Partially supported by a grant from Capes/Brazil}}
\address{Centro Brasileiro de Pesquisas F\ii sicas\\
Rua Xavier Sigaud 150, 22290-180, Rio de Janeiro, RJ, Brasil}
\date{\today}
\maketitle
\vskip 1.0 true cm
\centerline{$^{\dagger}$For the D\O\ Collaboration}
\vskip 1.0 true cm
\begin{abstract}
We present the first results of detector R \& D done for the proposed
Forward Proton Detector at D\O. From a menu of options we have
chosen a scintillating fiber based detector with multi-anode
photomultiplier readout.

\end{abstract}

\section{Physics Motivation}

	The interest in diffractive scattering has been growing since the 
observation of diffractive production of jets, by the UA8 collaboration\cite{UA8}.  
This observation came in support of the recently introduced field of hard 
diffraction, in which the diffractive scattering is accompanied by high 
transverse momentum objects\cite{IS}. 

In principle such kind of events could be treated in terms of parton 
language, defining a partonic structure for the exchanged object.
Diffractive events can also be characterized by rapidity gaps, regions 
of the phase space with no particles. This is due to the colorless nature of the 
exchanged object. Diffractive kinematics are summarized in Fig. 1.

The experimental difficulty in observing these events is that the 
scattered proton tends to remain in the beam pipe and can not be detected 
using the typical collider central detectors. For that it is necessary to add 
special forward particle detectors to the central assembly at very small 
angles, or large distances from the collision point.
	
Although rapidity gap techniques give some information on the 
diffractive event, only the addition of a forward particle detector would 
allow access to the full kinematics of the scattered particle.
	
Other motivation for adding a forward particle detector follows:

\begin{itemize}
\item The Interest in Hard Diffraction has been growing dramatically due to 
recent results from HERA  (H1, ZEUS) and Tevatron (CDF, D\O).

\item For the next 10 years much of this physics can only be done at the 
Tevatron collider. Diffractive systems with masses greater than 450 $GeV/c^{2}$ 
can be produced at Tevatron compared to only 70 $GeV/c^{2}$ at HERA.

\item Some processes like diffractive dijet production are well established, but 
not accurately measured.

\item Other processes like hard double pomeron exchange are not yet observed, 
but are likely to exist.

\item There are chances for discovery of new physics, like centauros, glueballs, 
etc.
\item Since there is little data on elastic scattering except at small t, such a
 detector could extend our knowledge to the high portion of the |t| spectrum.

\end{itemize}

\begin{figure}
\centerline{\psfig{figure=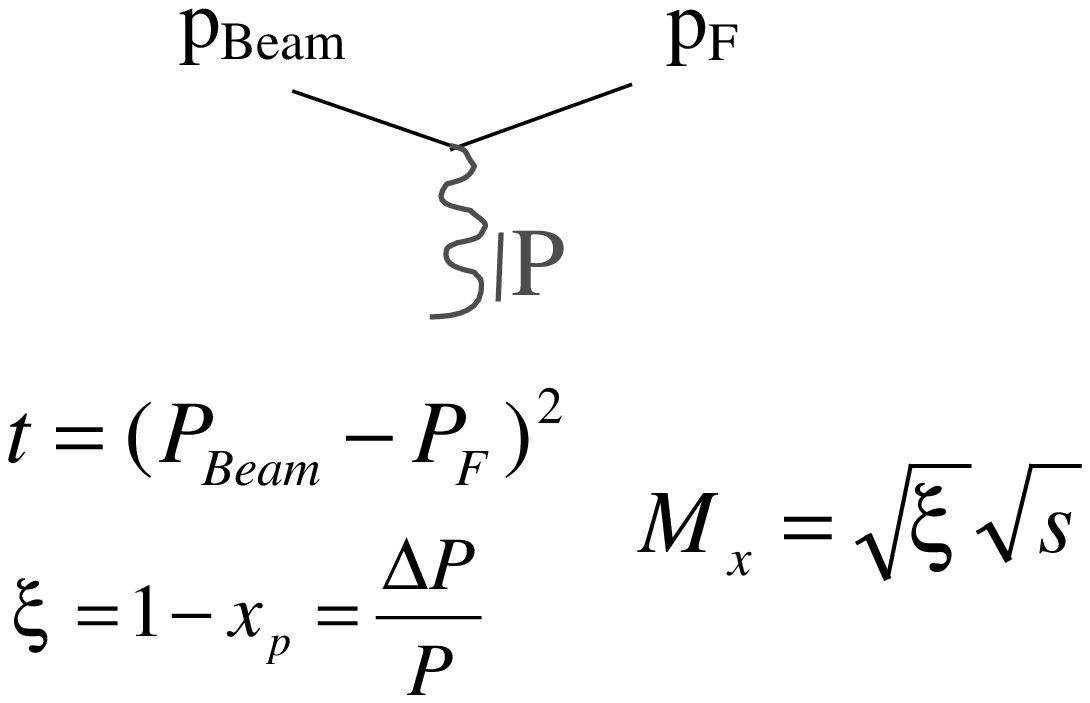,width=10cm}}
\caption{Kinematic variables in diffractive scattering. Here t corresponds to
the four-momentum transfer, $\sqrt{s}$ 
is the center of mass energy, $x_p$ is the fractional momentum carried by the
scattered proton, and $M_x$ is the 
diffractive mass.}
\end{figure}

\section{The Forward Proton Detector}

The Forward Proton Detector\cite{FPD} is a series of momentum spectrometers that 
make use of machine magnets in conjunction with position detectors along 
the beam line in order to determine the kinematic variables (t and $\xi$) of the 
scattered protons and anti-protons. The position detectors have to operate a 
few mm from the beam position and have to be moved away during the 
machine injection time. Special devices called ``Roman pots'' are designed 
to house the position detectors allowing for remotely controlled movement 
with good accuracy. The Roman pots for the Forward Proton Detector are 
described in a separate work presented at this conference\cite{POT}. The greatest 
challenge for the use of these devices in our case, is the limited space 
available around the D\O\ interaction region. 

Figure 2 shows a schematic representation of the location of the Roman Pots 
with respect to the D\O\ collision point. Spectrometers are labeled with respect 
to the analyzing magnet used. While the dipole spectrometer has detectors 
only in the horizontal direction, quadrupole spectrometers have detectors in 
both horizontal and vertical directions in order to maximize the acceptance.

This layout requires a total of 18 position detectors for both dipole and 
quadrupole spectrometers. It gives us the ability to trigger on both scattered 
protons and anti-protons while the hard scattering products are measured 
using the full D\O\ detector.

\section{Detector Goals}

In order to characterize the hard diffractive events we need to be able to 
measure its kinematic variables to a good precision. However, external 
factors like the beam dispersion, uncertainty in beam position, multiple 
scattering and Roman Pot positioning limit the attainable resolution. Hence 
a point resolution of about 100$\mu$m is sufficient for the Forward Proton
Detector. Some 
desirable characteristics for this detector are listed bellow:

\begin{itemize}

\item Overall efficiency close to 100\%

\item Modest radiation hardness. The detector will be operating at 
approximately 8$\sigma$ ($\sigma$ is the beam standard deviation) from the beam axis, so 
the expected radiation dose for one year of running is only 0.03 Mrad.

\item High Rate capability. The detector needs to be live at every beam crossing.

\item Low sensitivity to accelerator background, in particular particle showering 
along the beam pipe or magnets.

\item Small dead area closer to the beam. This is particularly important since the 
protons are scattered at very low angles, so the detector acceptance is 
heavily dependent on its position relative to the beam axis.

\end{itemize}
	
\begin{figure}
\centerline{\psfig{figure=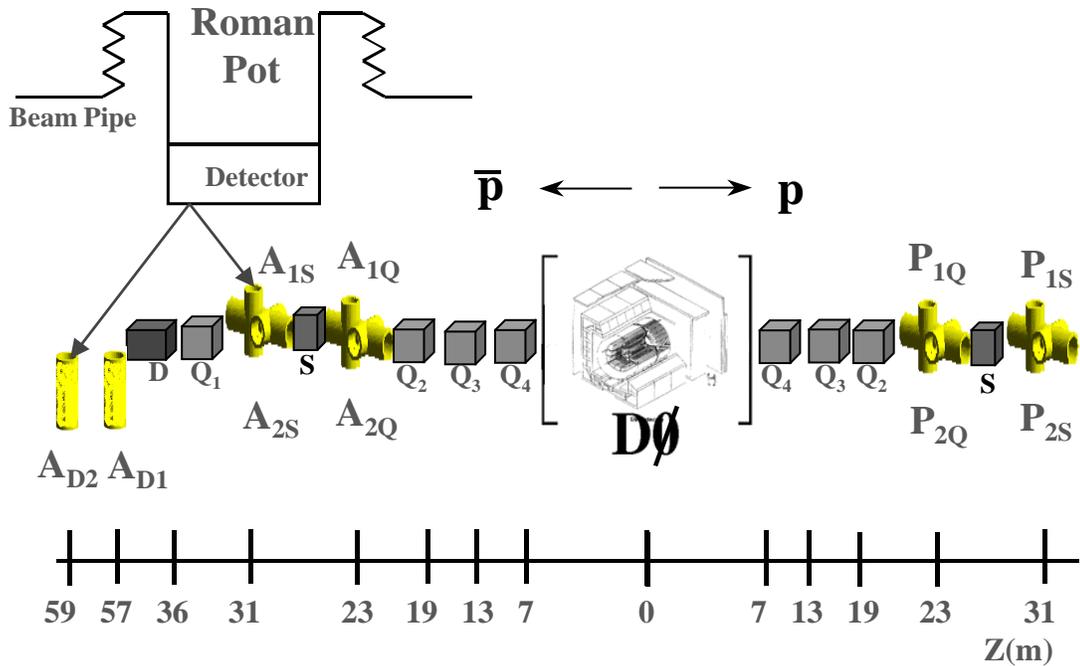,width=20cm}}
\caption{Schematic placement of the position detectors around the D\O\
collision hall. The labels A and P 
represent the anti-proton and proton side detectors respectively, while Q, S and
D label the quadrupole, 
separator, and dipole magnets.}
\end{figure}

\section{Detector Options}

We have considered several options for the position detector, some of 
which are listed bellow:

\begin{itemize}

\item Silicon Microstrip Detector
\item Microstrip Gas Chamber (MSGC)
\item Scintillating Fiber Detector

\end{itemize}

In Table I we present some of the advantages and disadvantages of each of 
these detector options.

\begin{table}
\caption{Detector Options for the FPD}
\begin{tabular}{l|l|l}
Detector & Pro & Con\\\hline
Silicon  & Superior   & Dead Area\\
         & Resolution & Trigger\\
MSGC     & Good       & Dead Area\\
         & Resolution &          \\
Fiber    & Adequate   & Light   \\
         & Resolution & Output\\\hline

\end{tabular}
\end{table}

	Silicon microstrip detectors have very good resolution and can operate 
at high rates, however they have a high cost/channel and problems with 
dead area at the bottom of the pot (1mm), where acceptance is crucial. Also 
the readout time for this detector is not fast enough to use it in a Level 1 
trigger, which we find necessary in order to reduce the event rates to an 
acceptable level.

	MSGCs suffer from an even worse dead area problem as compared to 
silicon detectors. Reducing this dead area could be possible, but it would 
involve considerable efforts in Research and Development (R \& D). 

Scintillating Fiber Detectors do not present the dead area problem and can 
be read out using fast devices like photomultiplier tubes or photodiodes. A 
drawback of these detectors is the small light yield obtained from fibers that
have a small enough diameter to give the desired position resolution. 
This can be seen from equation (1), which 
expresses the light yield as a function of several parameters, including 
the length of the fiber material traversed by the particle.

\centerline{\psfig{figure=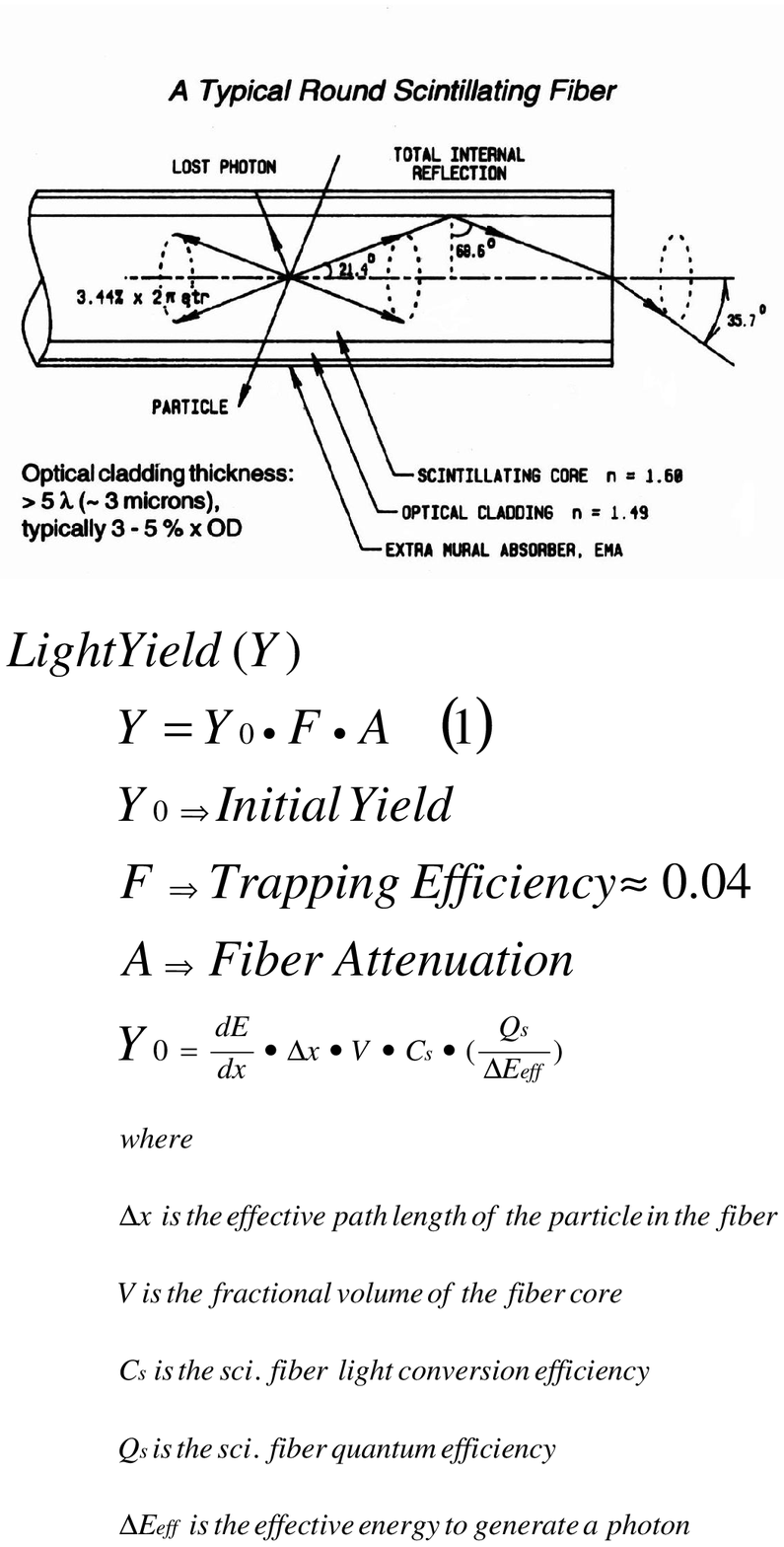,width=15cm}}
\vspace{-0.5cm}

	This problem can be overcomed by the use of very efficient photon 
detection devices, like the Visible Light Photon Counter (VLPC)\cite{VLPC}, the 
Avalanche Photo Diode (APD), the Image Intensifier with CCD readout, 
and the Multi Anode Photomultiplier (MAPMT) among others. A 
comparison of the Quantum Efficiency (light to charge conversion) between 
these devices is given bellow.

\begin{center}
\begin{tabular}{c|c}

Device & Quantum Efficiency\\\hline
 VLPC & ($\approx$ 80\%)\\
 APD  & ($\approx$ 70\%)\\
 Image Intensifier + CCD  & ($\approx$ 20\%)\\
 MAPMT & ($\approx$ 20\%)\\\hline

\end{tabular}
\end{center}

	APDs are still in a development stage, especially for our application, 
which requires multiple channels in a single chip. Image Intensifiers with 
CCD readout cannot operate at the high rate needed by our detector. 
VLPCs, which were developed within the D\O\ collaboration, are by far the 
best option in terms of quantum efficiency, but the cost and complications 
of the cryogenics associated with it led us to the far simpler MAPMT. These 
devices have a reasonable sensitivity, high gain and can be operated at very 
high rates. We have chosen to use Hamamatsu H6568 16 channels 
MAPMT, which has single photoelectron (PE) sensitivity and good gain stability 
between channels.

	Within the fiber framework we have investigated several options as 
detection elements:

\begin{itemize}

\item  A thin piece of scintillator tile connected to clear fibers for readout.
\item  A thin piece of scintillator tile connected to wavelength shifting fibers for 
readout.
\item  Scintillating fiber bundle straight from the detector area to the MAPMT.
\item  Scintillating fiber bundle connected to clear fiber for readout.

\end{itemize}

	We have decided to use square fibers as opposed to round due to the fact 
that the former type gives an increase of about 20\% in the light output. We 
use 0.8mm square fibers, giving about 80$\mu$m point resolution, which is close 
to the desired value. The fibers were bundled in groups of four, since earlier 
studies have indicated that such an arrangement would give around 10 
photoelectrons. In all cases one end of the detector element was aluminized 
to increase the light yield.
	Figure 3 shows a schematic drawing of the test setup used to compare 
different detection elements. All tests were done using $^{90}_{38}Sr$ and $^{106}_{44}Ru$ 
radioactive sources collimated to the fiber dimension. A single channel of the 
Hamamatsu H6568 16 channel MAPMT read out the four fiber bundles. 
Single PE calibration was done using a short pulse through a blue LED.

\begin{figure}
\vspace{1.5cm}
\centerline{\psfig{figure=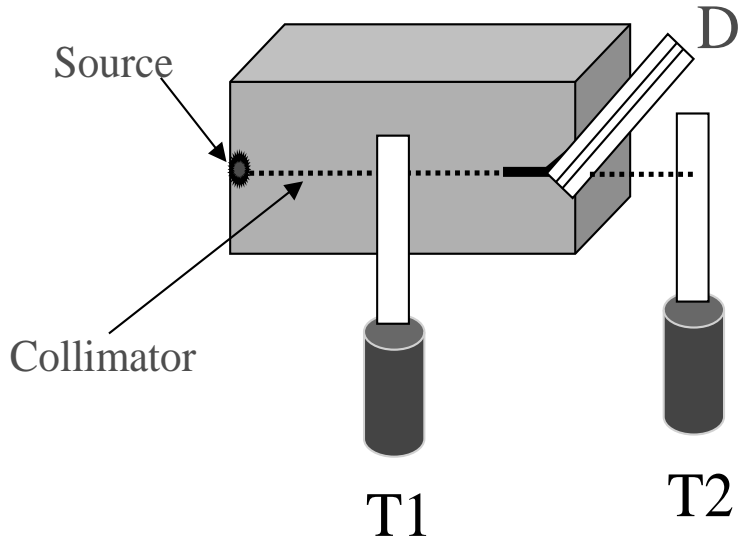,height=8cm,width=15cm}}
\caption{Source setup for testing FPD detector elements (D).  T1 and T2 are trigger scintillators. D is readout 
by a MAPMT (see text).}
\end{figure}

	Results show a clear advantage in using scintillating fibers in 
comparison to a scintillator tile. This could be due to a possible lower 
trapping efficiency for the thin tile. We have also noticed that the thin tiles 
are very fragile, often developing several fractures even being handled with 
extreme care. Figures 4, 5 and 6 show the integrated charge for each of the 
options tested.

\begin{figure}
\vspace{-1.5cm}
\centerline{\psfig{figure=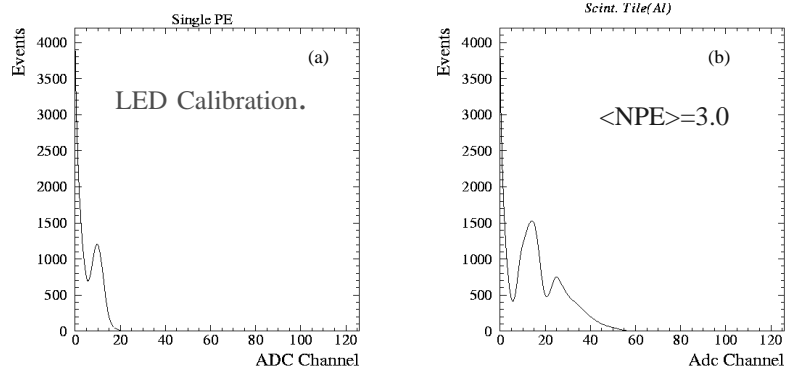,width=15cm}}
\caption{Integrated charge from MAPMT readout of detector elements. (a) shows the LED calibration of the 
single photoelectron peak. (b) is the charge distribution for a thin scintillator tile plus clear fiber as 
waveguide.}
\end{figure}

Figure 5 (b) shows the net effect of having the aluminized end of the detector 
cut at a $45^{o}$ angle. Having the fiber planes oriented at this angle (U, V planes 
instead of X, Y) would decrease the radius for the fiber bending, resulting in 
a more compact Roman Pot with the additional advantage of avoiding complicated 
pressure compensating mechanisms for this device. We can see that the 
effect of plane orientation is minimal.

\begin{figure}
\centerline{\psfig{figure=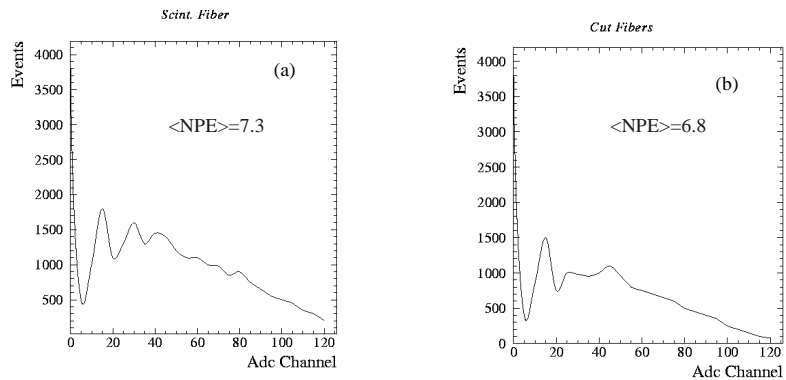,width=15cm}}
\caption{Integrated charge from MAPMT readout of detector elements. (a) shows
the distribution from four  
0.8mm thick by 50cm long scintillating fibers aluminized at one end, and (b) shows the same arrangement but 
with the aluminized end cut at a $45^{o}$ angle with respect to the fiber axis.}
\end{figure}

\begin{figure}
\centerline{\psfig{figure=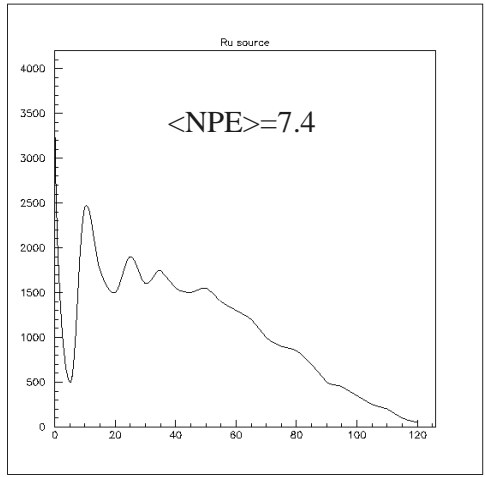,width=15cm}}
\vspace{-0.5cm}
\caption{Integrated charge distribution from MAPMT readout of a detector element consisting of 4 
scintillating fibers 0.8mm thick by 5cm long cut at a $45^{o}$ angle in one side and glued to 50cm long clear fibers 
as a waveguide.}
\end{figure}

	The setup using scintillating fibers spliced to clear fibers was chosen 
bearing in mind the effects of the halo background in the scintillating fibers 
alone, and also a slight decrease in the optical cross-talk with this setup. A 
comparison between Fig.6 and Fig.5 (b) shows that the scintillating plus 
clear fiber combination actually improves the light output due to the shorter 
attenuation length of the smaller scintillating fiber.
	Preliminary efficiency measurements done using the chosen setup 
range from 90-95\%. This measurement does still have a residual photon 
background from the radioactive source and we are working on a cosmic ray 
setup to solve this problem. We have also detected a dependence of this 
result on the cutting, mirroring and splicing procedures, which have to be 
monitored carefully.
	The final detector configuration is shown in Fig. 7. There will be 6 
planes in 3 views (U, V and X) in order to minimize ghost hit problems. The 
fibers will be mounted on a thin plastic frame, amounting to 112 channels 
per detector and a total of 2016 channels in all spectrometers.
	Each channel corresponds to 4 scintillating fibers, which fits nicely in 
the 4cm x 4cm-pixel size of the H6568 MAPMT. A total of 7 MAPMT will 
be needed for each detector, which represents the major component in the 
detector cost.
	This detector will allow us to make high-statistics and precise 
measurements of hard as well as inclusive diffractive processes, studying the 
dependence of these processes on kinematic variables.
	The combination of that with measurements in the central D\O\ 
detector will allow us to distinguish between different models for the 
pomeron structure.

\section{Conclusions}
	
	After considering several alternatives a scintillating fiber based 
detector with MAPMT readout was chosen as the best option for the FPD, 
considering all the aspects of acceptance, costs, radiation hardness and 
trigger capabilities.
	A prototype detector is being built at FNAL for resolution studies, and 
will be tested using a cosmic ray and, if available, a test beam setup.
	The FPD will add to the already existing diffractive physics program 
of D\O, being completely integrated to the existing detector.
	
	The field of Hard Diffraction is in clear need of more precise data for 
a better understanding of the diffractive phenomena, and the FPD will 
certainly help to achieve this goal.
	
	The arrangement of quadrupole and dipole spectrometers that we have 
chosen will extend the kinematic range of previous detectors, helps us in a 
better understanding of alignment and background issues, and could also be 
used for luminosity measurements.

\begin{figure}
\vspace{-2.0cm}
\centerline{\psfig{figure=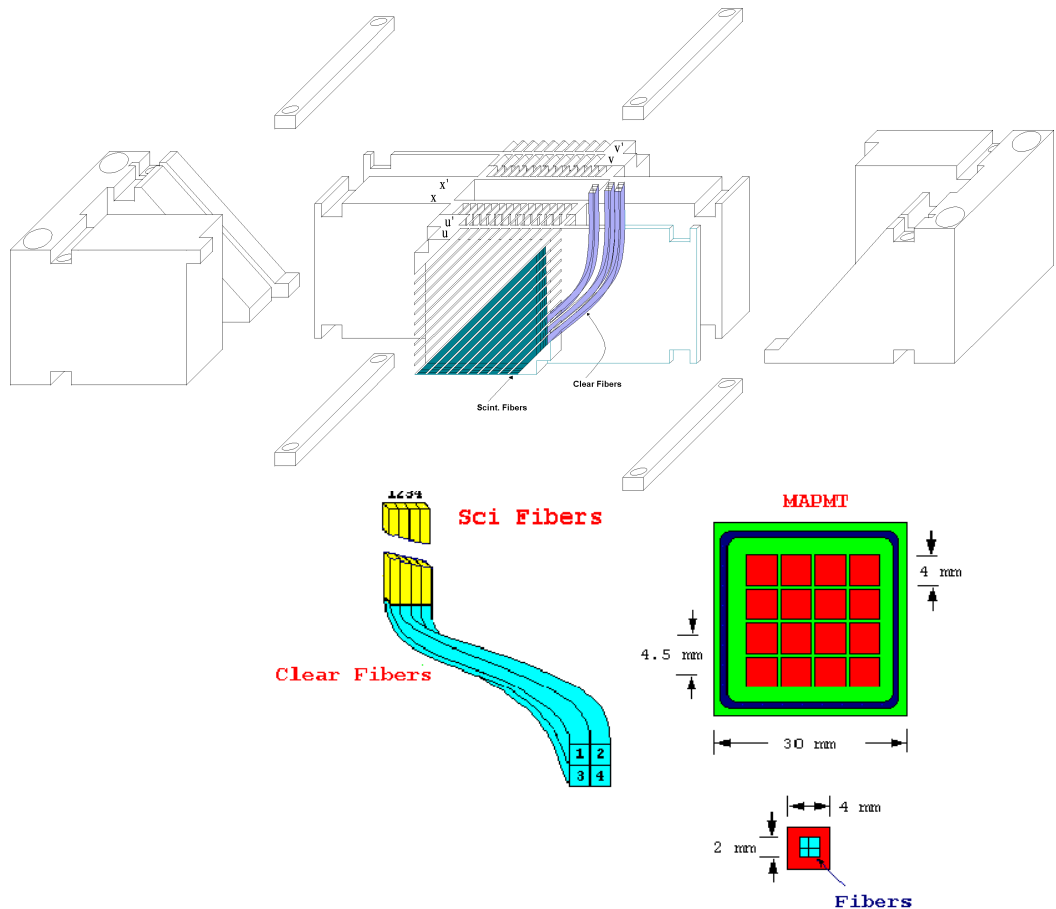,width=18cm}}
\vspace{-4.0cm}
\caption{Proposed design for each FPD detector. Also shown are the frames for fiber alignment and the fiber 
to photomultiplier interface.}
\end{figure}

\end{document}